# Productive Dialog During Collaborative Problem Solving

Robert G.M. Hausmann, Brett van de Sande, Carla van de Sande, and Kurt VanLehn
University of Pittsburgh, 3939 O'Hara Street, Pittsburgh, PA, 15260-5179
[bobhaus, bvds, cav10]@pitt.edu, vanlehn@cs.pitt.edu

**Abstract:** Collaboration is an important problem-solving skill; however, novice collaboration generally benefits from some kind of support. One possibility for supporting productive conversations between collaborators is to encourage pairs of students to provide explanations for their problem-solving steps. To test this possibility, we contrasted individuals who were instructed to self-explain problem-solving steps with dyads who were instructed to jointly explain problem-solving steps in the context of an intelligent tutoring system (ITS). The results suggest that collaboratively developed explanations prompted students to remediate their errors in dialog, as opposed to relying on the ITS for assistance, which is provided in the form of on-demand hints. The paper concludes with a discussion about implications for combining proven learning interventions.

## Introduction

As is evident to those who live and work in societies with advanced technologies, the world is not only becoming a smaller place, but the demands for collaboration are expanding across disciplinary (Schunn, Crowley, & Okada, 1998) and geographic boundaries (Friedman, 2006). Individuals are finding themselves collaborating in new ways that have been made possible by recent advances in high-speed networks and digital forms of communication. For individuals to stay competitive on a global scale, they need to develop their collaborative skills. The field of the learning sciences is uniquely positioned to provide recommendations for how to best optimize those collaborative skills.

In the paper that follows, we attempt to develop the following argument. First, it is evident from the collaborative problem-solving literature that, when done "naturally," collaboration is not much more effective for learning gains than solo problem solving (Hill, 1982). Attempts to optimize collaborative learning have included various scripting manipulations that increase learning gains (Rummel & Spada, 2005). However, novices generally do not use these effective modes of interaction to communicate their ideas; therefore, the interactions must be taught. Moreover, attempts to use computers to elicit improved collaboration via scripts have floundered because of their inability to understand natural language (Soller, 2004). The research problem addressed in the current paper is how to use computers to help increase learning during collaboration.

Toward that end, the paper is organized into the following sections. First, we will highlight two effective learning situations: self-explanation and peer collaboration. Then we will introduce an intelligent tutoring system for physics, called Andes, which has also been shown to increase individual learning. After the background for the study has been presented, we will report on an experiment that contrasted self-explaining with peer explanation in the context of using the Andes physics tutor. Finally, we will conclude with a discussion about leveraging the impact of various learning interventions.

### Learning Alone: Self-explaining Worked-out Examples

When enrolled in a course like physics, much of the learning that takes place outside the classroom is done individually. That is, students are generally responsible for learning the course material from a textbook, which often contains worked-out examples. On first inspection, worked-out examples tend to be fairly impoverished, in the sense that they typically omit information that needs to be supplied by the learner (Chi & Bassok, 1989). While examples may exclude some information, students prefer to learn from them, especially during the initial acquisition of a skill (Pirolli & Anderson, 1985). How do students learn from incomplete worked-out examples? One hypothesis is that students attempt to explain the examples, line-by-line, to themselves (Chi, Bassok, Lewis, Reimann, & Glaser, 1989). This study strategy goes by the name of *the self-explanation effect* (Chi, 2000).

Self-explaining is a constructive learning activity because the student is actively trying to make sense of the material from his or her own background knowledge. For instance, consider the following monolog from a student in a second-semester physics class (Hausmann & VanLehn, 2007). In this experiment, the student was asked to study an example *after* solving an isomorphic problem. The example was broken down into problem-solving steps, which were related to the motion of a charged particle in a region of a uniform electric field. In this episode, the student had just watched a video-based example of a step where the solver drew a force vector in the opposite direction of the electric field vector (as per the vector equation $\mathbf{F} = q\mathbf{E}$, where $\mathbf{F}$ is the force due to the electric field; $q$ is the charge on the particle, which is negative in this instance; and $\mathbf{E}$ is the electric field). The student is attempting to make sense of this step (see Table 1).

Table 1: An example of a self-explanation (SE) episode.

| Line | Code | Statement |
|---|---|---|
| 1 | Paraphrase | Up until this step, we've just been putting in information that was given. But now we have to apply concepts of physics. Um, we take from paragraph we know that... or what we're trying to find is the force of the electric field. So you go over to your menu on the left-hand side. And click the force vector ((*inaudible*)). Click and drag. Dialog box comes up. And you... the force is on the particle that we identified earlier due to an unspecified force. We know that it's electric. Um, we need to put what angle it is at. |
| 2 | SE: meta-cognitive | This is something I don't fully understand yet, |
| 3 | SE: justification | but they said you had to add whatever it was... two twenty plus one eighty to get two oh two. I guess that means it's in the opposite direction of the electric field. |

The episode opens with the student paraphrasing the material from the example (line: 1). This is nearly a verbatim representation of the step articulated in the video. In the second line, the student expresses some uncertainty about the step. This is coded as a "meta-cognitive" statement because the student is reflecting on her current state of understanding. In the last line, the student unpacks the calculation for determining the angle of the vector (i.e., 22 deg + 180 deg = 202 deg) and connects the addition of 180 deg to the concept of the force in the opposite direction of the given electric field.

This episode is typical of self-explaining because it begins with a paraphrase (Magliano, Wiemer-Hastings, Millis, Munoz, & McNamara, 2002), makes an identification of what is confusing, and then the confusion is remediated by supplying an additional piece of information (Chi, 2000; Chi & Bassok, 1989; Chi et al., 1989). The design of the experiment was also consistent with the observation that alternating between worked examples and solving problems is an effective method for acquiring a new cognitive skill, such as Lisp programming (Trafton & Reiser, 1993). Moreover, focusing self-explanations on individual steps of a solution has also been shown to be particularly effective in learning a meta-cognitive strategy, such as solving problems in Geometry (Aleven & Koedinger, 2002).

## Learning Together: Collaborative Peer Learning

While generating explanations on one's own has been shown to be effective in a number of different settings and domains, instructors also insist that students must learn to collaborate on projects as well. Peer collaboration can be an effective learning situation when it meets certain preconditions. For instance, peer collaboration seems to provide optimal learning outcomes when the students' interactions are scripted or scaffolded. For example, in a simulated clinical task, students were asked to collaboratively develop a diagnosis and therapy plan for a psychological disorder. Collaborators in a scripted condition produced better joint solutions than the dyads in an unscripted condition (Rummel & Spada, 2005). This result suggests that the default collaboration behaviors that many students use may not be fully optimal for solving problems.

Parallel evidence for this claim has also been found in a conceptual engineering task. Instead of scripting the dialog, students were given instructions at the beginning of the task to engage in elaborative dialogs. Participants in the control condition were given instructions on being responsive, but were not instructed to converse in any specific way. The results were fairly straight-forward. Students who were instructed to elaborate designed better optimized bridges and learned more deep knowledge than students in the control condition (Hausmann, 2006). In this particular study, the dialog was not heavily scripted. This suggests that effective collaboration can occur, even in lightly scripted learning situations.

## Andes: An Intelligent Tutoring System for Physics

Thus, self-explaining and peer collaboration are both effective learning situations. Their estimated effect sizes are $d$ = .74 – 1.12 for self-explaining (Chi et al., 1989; McNamara, 2004) and $d$ = .21 – .88 for peer collaboration (Johnson & Johnson, 1992; Slavin, 1990). The effectiveness of intelligent tutoring systems falls between these two estimates, somewhere in the neighborhood of $\sigma = 1.0$ (Anderson, Corbett, Koedinger, & Pelletier, 1995). The Andes[1] system, an intelligent tutoring system for physics, has also demonstrated comparable effect sizes (VanLehn et al., 2005). Andes has been in use at the U.S. Naval Academy since 1999. It covers nearly all of the topics (save thermodynamics and "modern physics") from a two-semester introductory physics course. One of the reasons for its effectiveness is "coached problem solving," in the sense that it offers instructional assistance at multiple levels of specificity. At the most general level, Andes provides instructional support in the form of "flag feedback" (see Fig. 1). Entries are flagged by turning correct entries green and incorrect entries red. The flags serve as unobtrusive indicators of progress as the student is solving the problem.

Andes also offers on-demand hints to remind students when they need to apply certain steps and how to get around impasses. The hints are designed to assist students to take the next correct problem-solving step. Often during problem solving, a student will complete all the steps that he knows, at which point he lacks the knowledge to take the next appropriate step. Andes supports the student in this situation by prompting him with successive levels of help. At the terminal level, Andes gives a bottom-out hint that explicitly tells the student what to do.

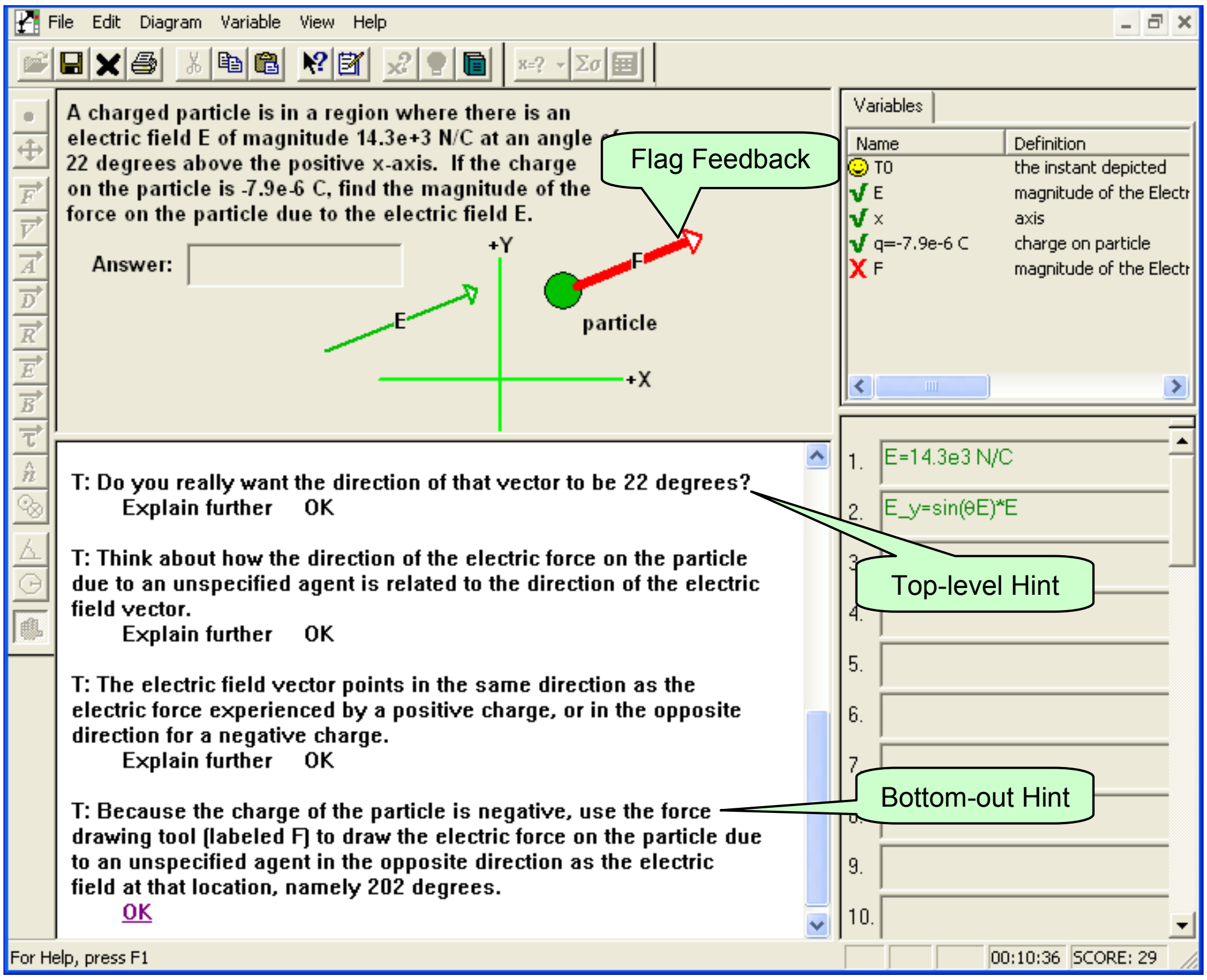

Figure 1. A screenshot of the Andes physics tutor, showing a partially solved problem. Each student entry turns red/green to indicate its correctness. On-demand hints are shown in the lower-left area.

## Method

In the previous sections, we reviewed three learning interventions that have been shown to work effectively with individuals: self-explanation of example steps (Aleven & Koedinger, 2002; Atkinson, Derry, Renkl, & Wortham, 2000; Chi et al., 1989; Renkl, 1997); peer collaboration (Stahl, 2006), and intelligent tutoring systems (Anderson et al., 1995; Mitrovic & Ohlsson, 1999; VanLehn et al., 2005). The hypothesis that we tested is as follows: If all of three aforementioned, effective learning situations are used as the context for collaboration (instead of other contexts, such as computer-supported collaborative learning), then collaboration will be more effective than individual self-explaining.

### Participants

Thirty-nine undergraduates ($N$ = 39), enrolled in a second semester physics course, were randomly assigned to one of two experimental conditions: self-explanation (individuals; $n$ = 11) or joint-explanation (dyads; $n$ = 14). Volunteers were recruited from several sections of a second semester physics course, which covered Electricity and Magnetism. Participants were recruited during the third week of the semester, with the intention that the experimental materials would coincide with their introduction in the actual physics course. The participants were paid $10 per hour. To ensure that the participants' motivation remained high during the entire two-hour session, they were offered an incentive of an additional $10 for doing well on the tests. All of the students received the bonus.

## Materials

The materials developed for this experiment were adapted from an earlier experiment (Hausmann & VanLehn, 2007). The domain selected for this experiment was electrodynamics with a focus on the definition of the electric field, which is expressed by the vector equation: $\mathbf{F} = q\mathbf{E}$. This particular topic is typically covered within the first few weeks of a second-semester physics course. Thus, it is an important concept for students to learn because it represents their first exposure to the idea that a field can exert a force on a body.

To instruct the participants, several materials were developed[2]. Four electrodynamics problems were created. These problems are representative of typical problems found in at the end of a chapter in a traditional physics textbook. The problems covered a variety of topics, including the definition of the electric field; Newton's first and second law, the weight law, and several kinematics equations. Each of the four problems was implemented in Andes. Andes was chosen because its design allowed for both the presentation of video-based examples, as well as coached problem solving (Conati & VanLehn, 2000). The first problem served as a warm-up problem because none of the students had any prior experience with the Andes user interface. In addition to the problems, three examples were created in collaboration with two physics instructors at the U.S. Naval Academy. The examples contained a voice-over narration of an expert solving the problems, and they were structured such that they were isomorphic to the immediately preceding problem.

## Procedure

To illustrate the procedure, consider a hypothetical participant in the self-explanation condition. The first activity was to watch a short, introductory video on the Andes user interface. Afterwards, she reads instructions on how to self-explain, including an example. She then solves the warm-up problem using Andes. During the problem solving, the student has access to the flag feedback, the hint sequences, and the Equation Cheat Sheet. Once the student submits a final answer, she opens the next example and watches an expert solution of an isomorphic problem. At the conclusion of each step of the video-based example, she is prompted to self-explain. The participant then verbally generates an explanation, which typically contains meta-cognitive statements, inferences, or questions she may have about the solution procedure. Once she is done with her self-explanation, she clicks a button to advance to the next step. Only the cover story and given values differ between the problem-solving and example problems. The student then alternates between solving problems and studying examples until all four problems are solved and all three examples are studied, or until two hours elapse. The participants in the joint-explanation condition followed the exact same procedure. The only difference was the instructions and prompts to jointly explain.

# Results

The results section is broken down into two parts. First, the data from the participants' problem-solving performance is presented. Second, an analysis of the dialog is discussed.

## Degree of Assistance During Problem-solving

As stated previously, Andes will always coach a student through a problem to its completion. This fact has two implications for the present study. First, solution accuracy will not discriminate between the two experimental conditions because, eventually, everyone solves the problem. Second, hint usage, especially bottom-out hints, are a compelling indicator of the degree of assistance needed during problem-solving. Students who rely on them generally lack a full understanding of the problem solution. Bottom-out hints were selected as the metric to gauge the quality of problem solving.

To assess the bottom-out hint usage, we counted the number of bottom-out hints per problem, then divided by the total number of student entries made for that problem. This metric is the rate of bottom-out hint usage. The self-explanation condition ($M = .16$, $SD = .17$) demonstrate a higher rate of bottom-out hint requests than the students in the joint-explanation condition ($M = .05$, $SD = .07$) (see Fig. 2). A one-way ANOVA was used to test the effect of condition on the total mean number of requested bottom-out hints, collapsing across all problems. There was a reliable effect of condition on the total number of bottom-out hint requests, $F(1, 23) = 4.896$, $p < .037$, $d = .93$. When a repeated measures ANOVA was used, all but one problem (Problem 2) were reliably different.

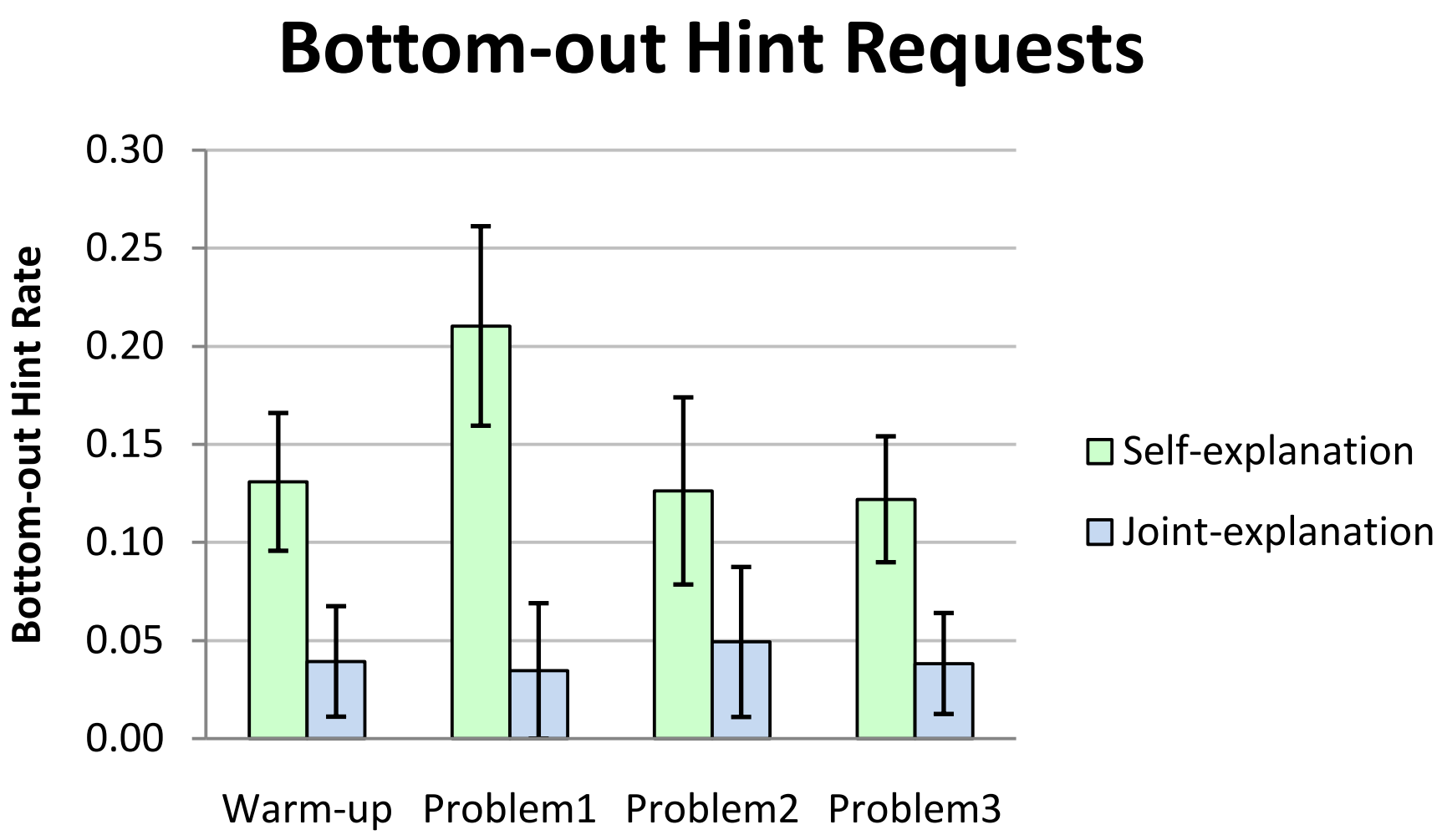

Figure 2. The mean (±SE) rate of bottom-out hint requests (number of hints/number of student entries) for each experimental condition.

In addition to the bottom-out hints, we analyzed the total number of on-demand hints received. Collapsing across all four problems and averaging over students per condition, the findings are entirely consistent with the bottom-out hint results. The individuals in the self-explanation condition ($M$ = 94.09, $SD$ = 51.92) requested nearly twice as many hints as the dyads in the joint-explanation condition ($M$ = 48.57, $SD$ = 39.59). When the pattern of means was tested with a one-way ANOVA, the results were both statistically and practically significant, $F(1, 23) = 6.20$, $p = .020$, $d = 1.05$. Both the hint and bottom-out hint results suggest that the dyads were able to fix their incorrect entries and impasses with less assistance from the Andes help system.

## An Analysis of the Collaborative Interactions

Why did the dyads ask for fewer hints? An analysis of the collaborative dialogs may offer an explanation of the disparate hint use. Instead of asking Andes for a bottom-out hint, dyads were able to interact with each other instead of the system. That is, dyads were able to draw upon the distributed expertise and extra cognitive resources to solve the problems with less explicit support. Consider the following exchange (see Table 2).

In this episode, the partners (Beth and Abby) attempted to apply the definition of the electric field ($\mathbf{F} = q\mathbf{E}$). They chose the correct principle to apply, but they chose the wrong *form* of the equation. That is, the equation $\mathbf{F} = q\mathbf{E}$ can be expressed in three ways. It can be expressed in terms of its components (i.e., F_x = qE_x & F_y = qE_y), as a vector equation (i.e., $\mathbf{F} = q\mathbf{E}$), or in terms of its magnitude (i.e., F = abs(q)E). In the present case, the magnitude form of the equation is considered the correct expression. Therefore, the equation calls for taking the absolute value of the charge because the sign on the change is irrelevant to a magnitude, which is always positive by definition. Beth and Abby are forgetting that they need to use the absolute value (written as "abs" in the transcript).

Table 2: An example of dyad exhibiting collaborative error remediation.

| Line | Speaker | Statement |
|---|---|---|
| Second Opportunity to Apply F = abs(q)E | | |
| 1 | Abby | And then, E equals |
| 2 | Beth | q times |
| 3 | Abby | Er, no. Fg divided by |
| 4 | Beth | Force is q times field. |
| 5 | Abby | A-huh. |
| 6 | Beth | So then field is...a force q times field |
| 7 | Abby | So it's q divided by |
| 8 | Beth | q divided by For-, yeah, there you go. [Types: E=F/q and Andes colors it red] |
| 9 | Abby | M'kay. |
| 10 | Beth | Er, is it the other way around? q divided by force? [Types: E=q/F. Andes colors it red & pops up a hint] |
| 11 | Abby | [Reads hint:] "Units are inconsistent." |
| 12 | Beth | I think it's Force divided by q because it's in Newtons per Coulomb. |
| 13 | Abby | Yeah. |
| 14 | Beth | So it was, it was, let's see what's wrong. |
| 15 | Abby | [Types: E=F/q] Hit enter. [Andes colors it red] |
| 16 | Beth | What's wrong?[Clicks on hint button, which displays, "Normally this equation is written using an absolute value."] |
| 17 | Both | Oh! |
| 18 | Abby | So, abs |
| 19 | Beth | Okay. |
| 20 | Abby | And solve for... |
| Third Opportunity to Apply F = abs(q)E | | |
| 21 | Abby | And then... |
| 22 | Beth | F equals q times E? |
| 23 | Beth | I remembered it this time. |
| 24 | Abby | Good job! |

In the first opportunity, Beth and Abby attempt to write the equation in terms of the electric field (i.e., E = F/q). However, when they write the equation, Andes flags the entry as incorrect (i.e., it turns it red) because they forgot to take the absolute value of the charge (q). They mistakenly interpret the feedback to suggest that their algebraic manipulation was incorrect (line: 10), and they attempt to fix it by swapping q for F. That brings up another error message that the units are inconsistent, which is an algebraic mistake (line: 11). This particular hint is a strong clue that they had it right the first time, so they retype their initial equation and ask for a hint (line: 14). The hint reminds them that they need to take the absolute value, and this hint is strong enough that they do not require the bottom-out hint. During the next opportunity to apply the same principle (lines: 21-24), they do so without error. This suggests that their impasse, plus the interaction between themselves and the top-level hint, was enough for them to strongly encode this knowledge component.

Another reason why dyads asked for fewer hints may be related to the way they processed the top-level hints. As stated previously, Andes offers hints in a graded fashion (see Fig. 1 for an example of a full hint sequence). The first hint is very general and is intended to cue the recall of an applicable step. If that does not help, then the student can ask for an addition hint that is more specific. Dyads may be in a better position to understand the hints because they can discuss them with a partner. To demonstrate this process, consider the following dialog (see Table 3).

Table 3: An example of a dyad making sense of an on-demand hint.

| Line | Speaker | Statement |
|---|---|---|
| 1 | Andes | Do you really want the direction of that vector to be 1 degrees?<br>**Explain Further OK** |
| 2 | Andes | Think about how the direction of the electric force on the particle due to the unspecified agent is related to the direction of the electric field vector.<br>**Explain Further OK** |
| 3 | Kip | Okay, well, if the...is it gonna be moving in the same direction as the field? Like if the particle is negative, or will it move opposite the field? |
| 4 | Rex | Um…it will move in the direction of the field. |
| 5 | Kip | Okay, so I guess. |
| 6 | Rex | Wait! No, it should move opposite, shouldn't it? |
| 7 | Kip | /I always forget/ |
| 8 | Rex | /The line's / always going to the negative. |
| 9 | Kip | Okay, so it's a-, |
| 10 | Rex | So it should be opposite, shouldn't it? |
| 11 | Kip | Okay. |

The episode begins with the dyad asking Andes for help on drawing the electric force vector (line: 1-2). They are trying to figure out the direction of the force vector. The correct direction of the force vector is in the opposite direction of the electric field vector because the charge on the particle is negative. The dyad considers the hint very seriously and attempts to draw a relationship between the particle and the direction of the electric field vector. Kip poses the problem as a choice between two conflicting options (line: 3). He also poses the question as motion of the charged particle, which is interesting because the problem does not explicitly mention motion. Instead, he uses motion as a way to reason about the forces acting on the particle. Rex incorrectly answers the question by deciding that the motion of the particle should be in the direction of the electric field (line: 4). He then correct himself (line: 6), and he says it should be opposite. He hedges a little, which serves to invite additional input ("shouldn't it?"). Kip does not offer much of a reply, but they jointly decide it should be opposite. Through this process, they did not request a bottom-out hint.

## Discussion

While self-explaining worked-out examples, peer collaboration, and learning by solving problems with an intelligent tutoring system are each effective methods for improving learning, very few studies have made explicit comparisons across interventions. Our particular interest was attempting to decrease the variance in outcomes for collaborative dyads. Prior literature on peer collaboration suggests that the outcomes can be highly variable (Dillenbourg, Baker, Blaye, & O'Malley, 1995), and in sporadic cases, individuals outperforming groups (Hill, 1982). One method for increasing collaborative outcomes is to provide some sort of structure or scripting to the interactions. In the present case, we chose to structure loosely the interactions in two ways. First, we provided instructions for guiding the interactions while studying examples, which we called *joint-explanation*. The second method for structuring the dialog was to use a step-based tutoring system.

The problem-solving outcomes of the dyads suggest they were able to solve the problems with less instructional guidance by the intelligent tutoring system than the individuals. They requested fewer on-demand hints and fewer bottom-out hints than the individuals. One reason why they may have requested less assistance from the system is due to the dialog between the partners. That is, they helped each other make sense of the top-level hints. This possibility was demonstrated in two dialog excerpts.

Another possible explanation is that both the examples and the ITS helped focus the students on problem-solving steps. To gain the full benefit of learning from example-studying, however, the steps also need to be derived and explained. The ITS required the students to enter steps, but did not require any justification for them. Therefore, it was the individuals' and dyads' responsibility to determine *why* they should take a given step. We hypothesized that the pairs more frequently engaged in the sense-making activity of asking why. The outcome of this sense-making activity is a focus on a shared goal (i.e., explaining a step; doing a step), which then improved their collaboration. The pairs were better able to stay synchronized on the same goals; insuring that they jointly produced visible progress through the steps, and generally focused on coordinated, joint work. Future work will focus on a complete analysis of the verbal protocols generated by the individuals and dyads to see if our hypotheses about the learning processes are supported.

In summary, the results from this experiment demonstrate large effect sizes, when proven learning situations are combined. Self-explaining worked-out examples, peer collaboration, and learning while solving problems with an intelligent tutoring system have each shown strong improvements over baseline learning conditions (i.e., textbooks). How we maximize learning, via the combination of effective learning strategies, will help us construct new learning environments that go beyond the independent contributions of each single learning strategy.

## Endnotes

(1) To download a free copy of Andes, please visit: http://www.andes.pitt.edu/
(2) For a complete copy of the materials and instructions used in this experiment, please contact the authors.

## Acknowledgments

This work was supported by the Pittsburgh Science of Learning Center, which is funded by the National Science Foundation award number SBE-0354420. The authors are deeply indebted to Robert Shelby and Donald Treacy for their assistance in developing the instructional materials used in this experiment.